\documentclass[10pt, twocolumn, nofootinbib,superscriptaddress]{revtex4-1}
\usepackage{amsmath,amssymb,amsfonts}
\usepackage{algorithmic}
\usepackage{graphicx}
\usepackage{textcomp}
\usepackage{xcolor}
\usepackage{ragged2e}
\usepackage{booktabs, makecell, tabularx}
\usepackage{lipsum}

\usepackage{gensymb}
\makeatletter
    \renewcommand\@make@capt@title[2]{%
     \@ifx@empty\float@link{\@firstofone}{\expandafter\href\expandafter{\float@link}}%
      {\textbf{#1}}\@caption@fignum@sep#2\quad}%
\makeatother
\makeatletter 
\renewcommand{\fnum@figure}{\textbf{Fig.~\thefigure}} 
\makeatother

\usepackage{xcolor}

\def\BibTeX{{\rm B\kern-.05em{\sc i\kern-.025em b}\kern-.08em
    T\kern-.1667em\lower.7ex\hbox{E}\kern-.125emX}}

\begin{document}

\author{Kaixuan Ye}
\affiliation{Nonlinear Nanophotonics Group, MESA+ Institute of Nanotechnology, University of Twente, Enschede, Netherlands}
\author{Akshay Keloth}
\affiliation{Nonlinear Nanophotonics Group, MESA+ Institute of Nanotechnology, University of Twente, Enschede, Netherlands}
\author{Yisbel E. Marin}
\affiliation{VTT Technical Research Centre of Finland, 02044 Espoo, Finland}
\author{Matteo Cherchi}
\affiliation{VTT Technical Research Centre of Finland, 02044 Espoo, Finland}
\author{Timo Aalto}
\affiliation{VTT Technical Research Centre of Finland, 02044 Espoo, Finland}
\author{David Marpaung}
\email{david.marpaung@utwente.nl}
\affiliation{Nonlinear Nanophotonics Group, MESA+ Institute of Nanotechnology, University of Twente, Enschede, Netherlands}

\date{\today}
\title{Stimulated Brillouin scattering in a non-suspended ultra-low-loss thick-SOI platform}

\begin{abstract}
Silicon photonics, with its CMOS compatibility and high integration density, has enabled a wide range of novel applications. Harnessing stimulated Brillouin scattering (SBS), an optomechanic interaction between optical and GHz acoustic waves, in silicon-on-insulator (SOI) platforms attracts great interests for its potential in narrow-linewidth lasers and microwave photonics. However, the poor optoacoustic overlap in silicon nanowires on conventional SOI platforms has previously restricted the observation of SBS signals to suspended silicon waveguide structures. In this work, we report, for the first time, the SBS response in a non-suspended  ultra-low-loss thick-SOI waveguide platform. The SBS process in this 3~$\mu$m thick SOI platform is enabled by a leaky acoustic mode that coexists with the optical mode in the waveguide core, resulting in enhanced optoacoustic overlap. We measured a Brillouin gain coefficient of 2.5~m$^{-1}$W$^{-1}$ and 1.9~m$^{-1}$W$^{-1}$ at 37.6~GHz for the rib and strip waveguide, respectively. This work paves the way for Brillouin-based applications in non-suspended ultra-low-loss silicon photonics systems.

\end{abstract}
\maketitle
\section{Introduction}

Stimulated Brillouin scattering (SBS) is an optomechanic interaction between optical and GHz acoustic waves \cite{Eggleton2019}. It features an ultra-narrow linewidth gain window on the order of tens of MHz, with a characteristic Brillouin frequency shift on the order of tens of GHz. Exploiting SBS on integrated photonics platforms has unlocked various key technologies, such as ultra-narrow linewidth lasers \cite{Gundavarapu2018,Otterstrom2018}, high-selectivity microwave photonic filters \cite{Botter2022,Garrett2023}, and low-noise microwave/mmwave signal generator \cite{Bai2020}. Despite its huge potential, to date, SBS has only been demonstrated in a limited number of material platforms, including chalcogenide \cite{Pant2011,Garrett2023}, silicon nitride \cite{Botter2022,Botter2023, Gyger2020, Gundavarapu2018}, silicon oxynitride \cite{Ye2023,Ye2024,Zerbib2023}, and thin-film lithium niobate \cite{Ye2023-tfln, Rodrigues2023,rodrigues2023onchip}. 

While all these platforms are promising, they are still in the early stages of development compared to the mature SOI platforms for silicon photonics, which has been widely employed in datacenter interconnects \cite{Shekhar2024, Rizzo2023}, microwave photonics \cite{Marpaung2019IntegratedPhotonics, Garrett2023, Munk2019}, LiDAR \cite{Zhang-lidar2022, Rogers2021},  and optical computing \cite{Shen2017, Huang2021}.

Harnessing SBS in SOI platforms has attracted continuing interest \cite{Otterstrom2018, Kittlaus2016, Lei2024, Zhang2022, Ruano2024}. However, due to the higher acoustic velocity in the silicon waveguide compared to the silica substrate, the acoustic wave tends to leak into the substrate in conventional nanowire SOI platforms. This results in poor optoacoustic overlap, which has hindered the observation of SBS signals.

\begin{figure}[t!]
\centering
\includegraphics[width=\linewidth]{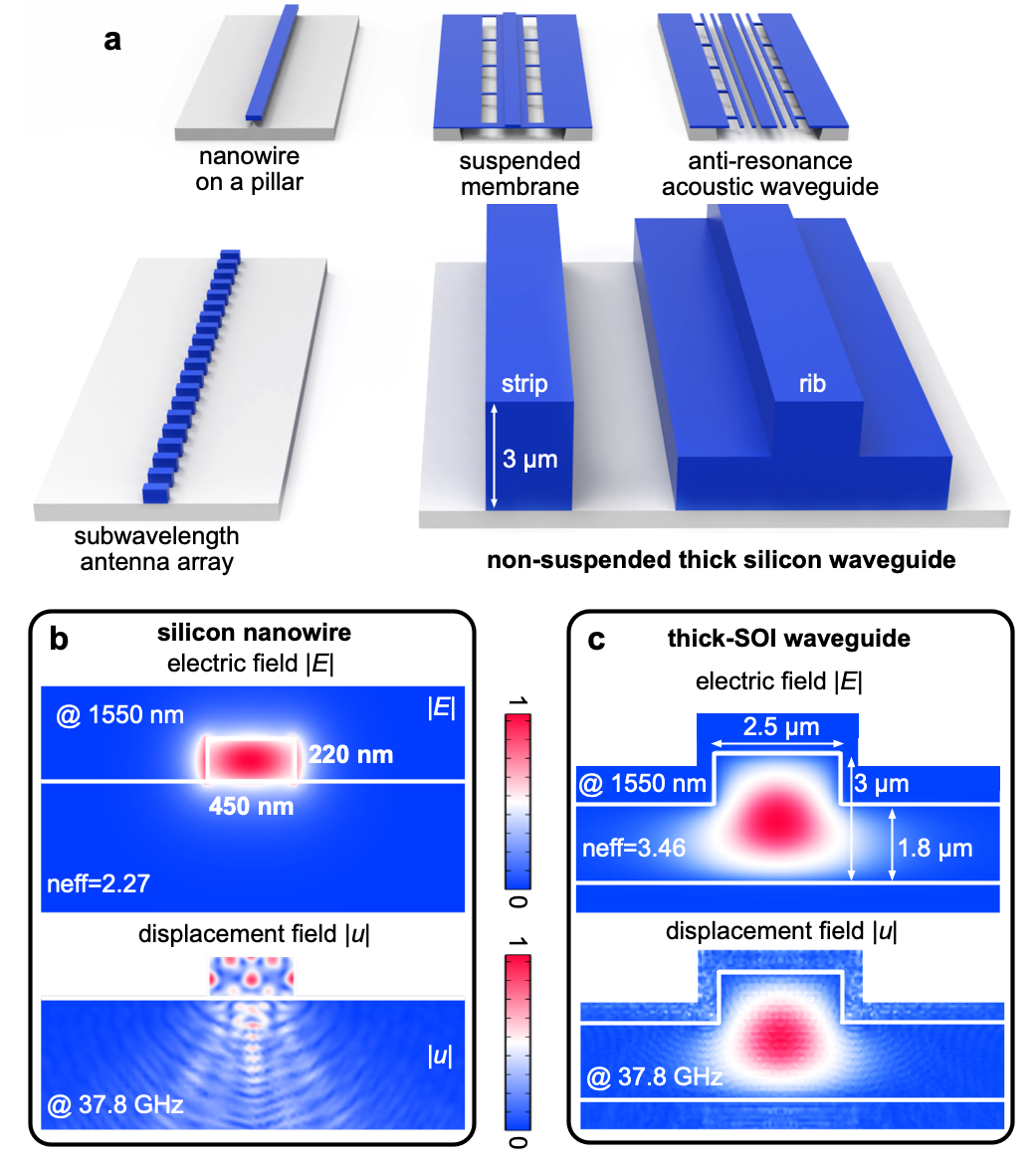}
\caption{\textbf{a} State-of-the-art SBS demonstrations in various SOI platforms (the dimensions are not to scale).  SBS in nanoscale SOI platforms typically requires suspended structures or subwavelength gratings; \textbf{b} Simulated electric and displacement field of a non-suspended SOI nanowire. The displacement field would significantly leak into the substrate, reducing the optoacoustic overlap.  \textbf{c} Simulated electric and displacement field of the non-suspended thick-SOI waveguide. Both fields can be well confined within the waveguide core area.}
\label{fig1}
\end{figure}

To mitigate acoustic wave leakage in nanowire SOI waveguides, different techniques have been proposed. Fig.~\ref{fig1}\textbf{a} presents state-of-the-art SBS demonstrations across various SOI platforms. In \cite{raphael2015}, a silicon nanowire was placed on a pillar to reduce the interface between the waveguide and the substrate. While effective, such a structure presents significant fabrication challenges and suffers from dimensionally induced inhomogeneous broadening. To alleviate these challenges, a membrane-suspended waveguide was introduced in \cite{Kittlaus2016}, where the Brillouin-active waveguide is supported by nanoscale tethers, enhancing the robustness of the structure. Such a structure also allows for independent control of both optical and acoustic modes. Building on this, a similar suspended structure with four reflective layers was proposed in \cite{Lei2024}. These reflective layers form Fabry-Pérot cavities for the acoustic wave, further enhancing the optoacoustic overlap within the waveguide core. Inspired by antenna array theory, an array of subwavelength antennas was employed in a non-suspended SOI platform to suppress acoustic leakage through destructive interference, as demonstrated in \cite{Zhang2022}.

These demonstrations with suspended nanowires can achieve a Brillouin gain coefficient of the order of hundreds to thousands m$^{-1}$W$^{-1}$, which are already useful for practical applications, such as Brillouin lasers \cite{Otterstrom2018} and microwave photonics filters \cite{Gertler2022}. However, the need for substrate removal demands high fabrication precision and limits the overall robustness of the device. Additionally, silicon nanowires typically exhibit propagation losses around 1 dB/cm, which hinders large-scale circuit integration. Furthermore, due to their small mode area, these structures are prone to large nonlinear loss from the two-photon absorption and the free-carrier absorption, limiting the power handling capabilities.

In this work, we investigate the feasibility of harnessing SBS in an ultra-low-loss and non-suspended thick-SOI platform. As opposed to silicon nanowires, thick-SOI can have significantly lower loss \cite{Aalto2019}, with reported propagation losses as low as 2.7~dB/m \cite{Marin2023}. Additionally, due to the larger mode area, thick-SOI waveguide experiences reduced nonlinear loss. We leverage these strengths of this technology to demonstrate, for the first time, the observation of SBS signals in a non-suspended thick-SOI platform. We measured Brillouin gain coefficient of 2.5~m$^{-1}$W$^{-1}$ and 1.9~m$^{-1}$W$^{-1}$ at 37.6~GHz from strip and rib waveguides. This work paves the way for Brillouin-based applications in a non-suspended ultra-low-loss
silicon platform.

\begin{figure}[t!]
\centering
\includegraphics[width=\linewidth]{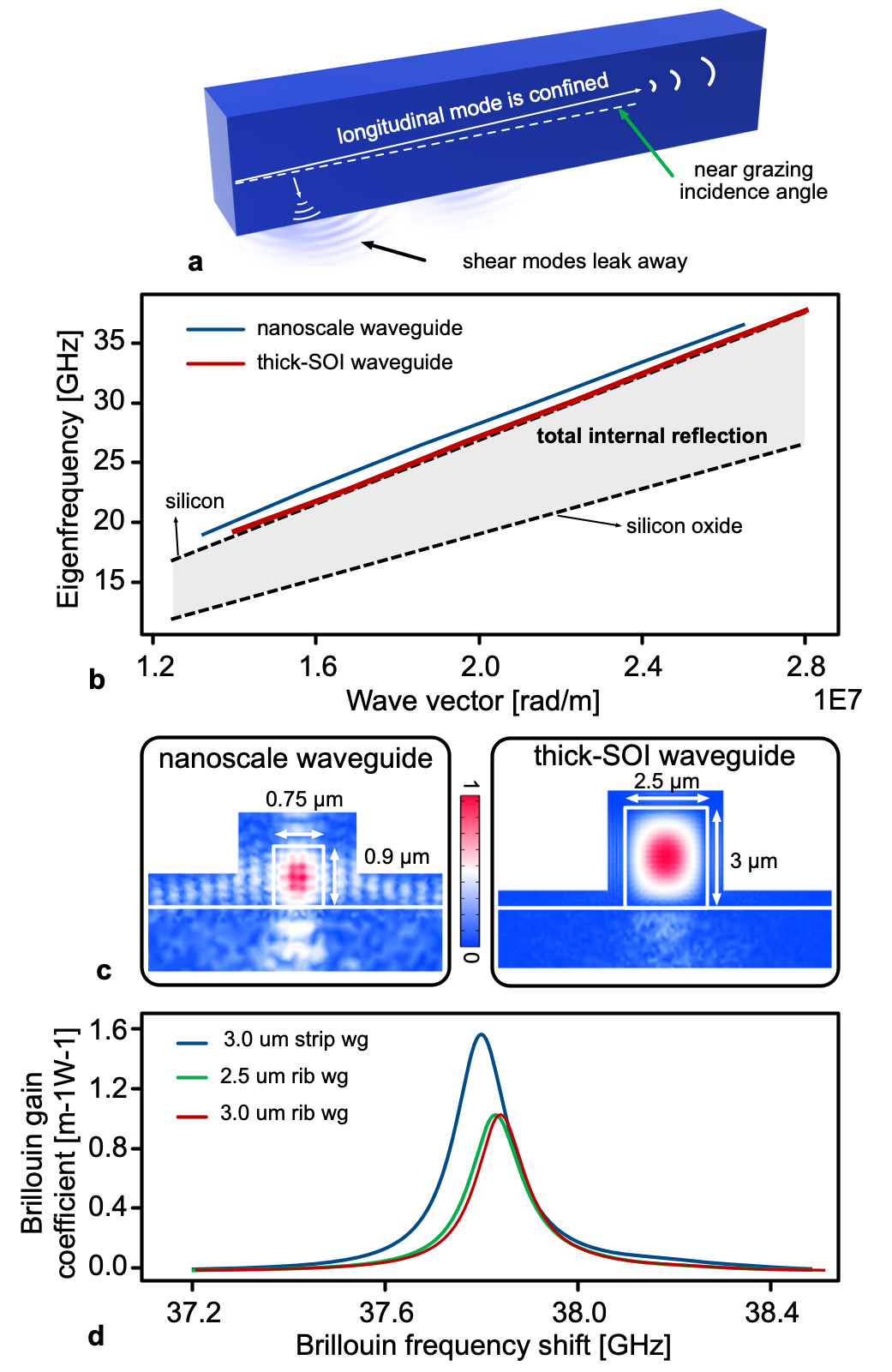}
\caption{\textbf{a} Explanations of acoustic wave confinement in thick-SOI waveguides. The longitudinal acoustic mode is confined due to the near-grazing incidence angle, while hybridized modes cause acoustic leakage because of their significant transverse components. \textbf{b} Dispersion diagram of acoustic modes in the thick-SOI waveguide and a smaller-dimension nanoscale SOI waveguide. \textbf{c} Simulated displacement fields in these two waveguides. \textbf{d} Simulated SBS responses of strip and rib waveguides in the thick-SOI platform.}
\label{fig2}
\end{figure}

\section{Acoustic wave confinement in thick-SOI waveguide}
SBS arises from the interaction between optical and acoustic waves, therefore, a large optoacoustic overlap is essential for an efficient SBS process. While the high refractive index of silicon (n = 3.45) ensures optical waveguiding through total internal reflection (TIR), this TIR condition does not apply to acoustic waves. Since the acoustic velocity in silicon, at 8500~m/s, is higher than that in silicon oxide (5960~m/s) \cite{Smith2016}, acoustic waves in conventional SOI nanowires tend to leak into the substrate, leading to a diminished optoacoustic overlap, as shown in Fig.~\ref{fig1}\textbf{b}.

Increasing the waveguide dimension can effectively mitigate acoustic wave leakage, even when the TIR condition breaks down \cite{Poulton2013}. As illustrated in Fig.~\ref{fig2}\textbf{a}, acoustic waves with dominant transverse components, such as shear-like waves, would leak at the silicon-silica interface because of the small incidence angle. However, acoustic waves that predominantly have longitudinal components would experience strong reflections at the waveguide-cladding interface due to the near-grazing incidence angle, analogous to light guiding in hollow-core fibers \cite{jonas2023}. Although energy still couples into the silica substrate and cladding, these leaky acoustic waves can be guided.

In the thick-SOI platform, benefiting from the large waveguide dimension, hybridization between longitudinal and shear acoustic modes becomes negligible. Consequently, such longitudinal-dominant acoustic mode exists in thick-SOI waveguides. As shown in Fig.~\ref{fig1}\textbf{c}, there is an acoustic mode coexisting with the optical mode in the same area, resulting in a large optoacoustic overlap.

Fig.~\ref{fig2}\textbf{b} plots the dispersion of acoustic modes in the thick-SOI waveguide and a smaller-dimension nanoscale SOI waveguide. Both acoustic modes lie outside the TIR zone, which is bounded by the dispersion of longitudinal modes in bulk silicon and silicon oxide. Nevertheless, the acoustic mode in the thick-SOI waveguide aligns more closely with the bulk silicon mode, whereas the mode in the smaller waveguide deviates further from the TIR zone, indicating stronger acoustic leakage. This is further highlighted by the comparison in Fig.\ref{fig2}\textbf{c}.

Unlike conventional modes confined under the TIR condition, leaky acoustic modes in thick-SOI waveguides exhibit complex eigenfrequencies, i.e., a limited mode lifetime. Despite this, this mode can still couple to the optical wave and contribute to the SBS process. If the lifetime of this mode is longer than the phonon lifetime in silicon, the interaction can be treated as a standard SBS process governed by a non-leaky acoustic mode. Otherwise, this leaky acoustic mode would broaden the SBS linewidth and reduce the peak gain value. 

Fig.~\ref{fig2}\textbf{d} presents the simulated SBS responses in the thick-SOI platform, with a Brillouin frequency shift of 37.79 GHz for the strip waveguides, slightly increasing to 37.83 GHz for the rib waveguides. Although the Brillouin gain coefficient in thick-SOI waveguides is lower than that demonstrated in suspended silicon nanowires, the gain coefficient observed in this work remains useful for microwave photonics applications \cite{Botter2022}. Additionally, the Brillouin frequency shift is much higher than what observed in other integrated platforms, underscoring the potential of the thick-SOI platform for pure millimeter-wave generation via the SBS process. All material properties used in our model are listed in Table 1.




\section{Non-suspended thick-SOI platform}
\begin{figure}[t!]
\centering
\includegraphics[width=\linewidth]{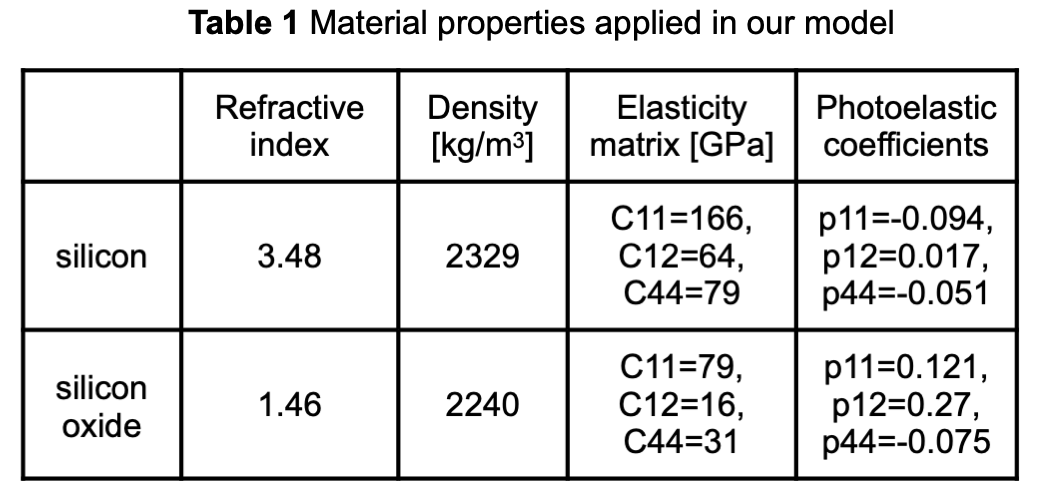}
\label{table}
\end{figure}

\begin{figure}[t!]
\centering
\includegraphics[width=\linewidth]{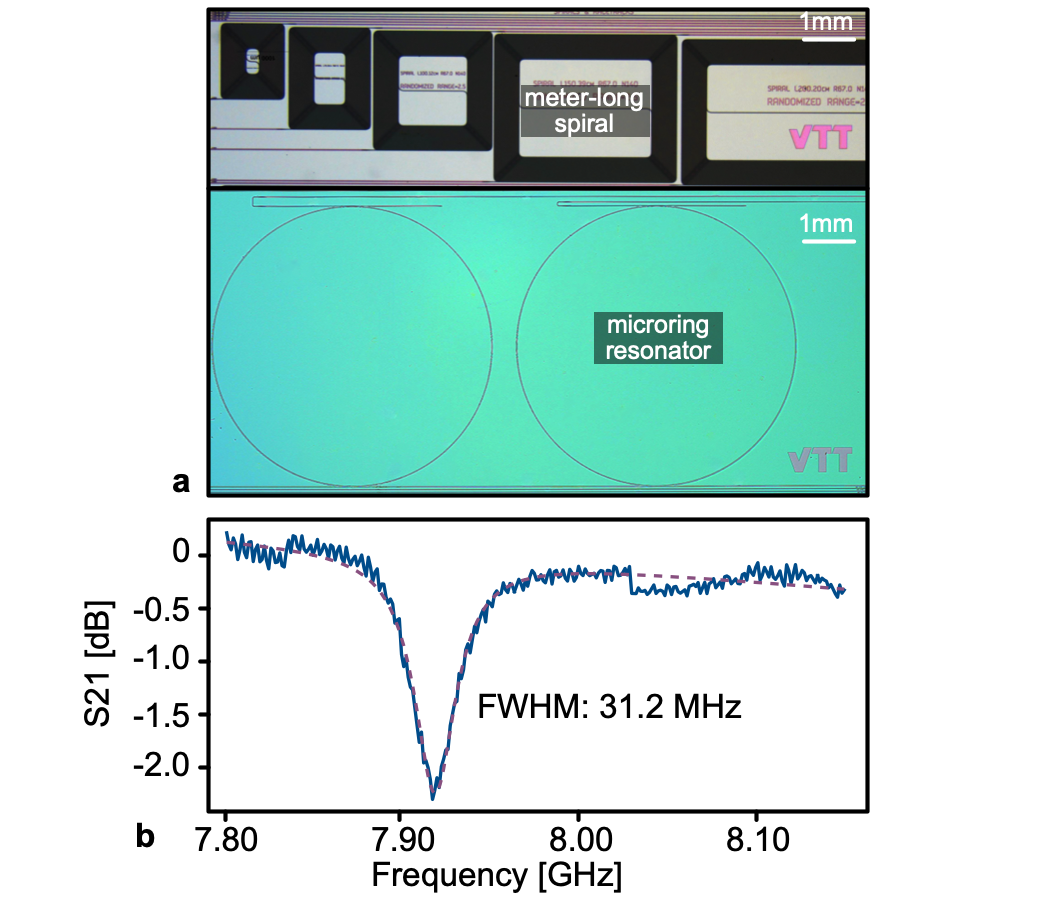}
\caption{\textbf{a} Microscope image of the thick-SOI waveguide sample containing meters-long spirals and add-drop microring resonators; \textbf{b} Resonance characterization of a microring resonator in the thick-SOI platform.}
\label{fig3}
\end{figure}

\begin{figure}[t!]
\centering
\includegraphics[width=\linewidth]{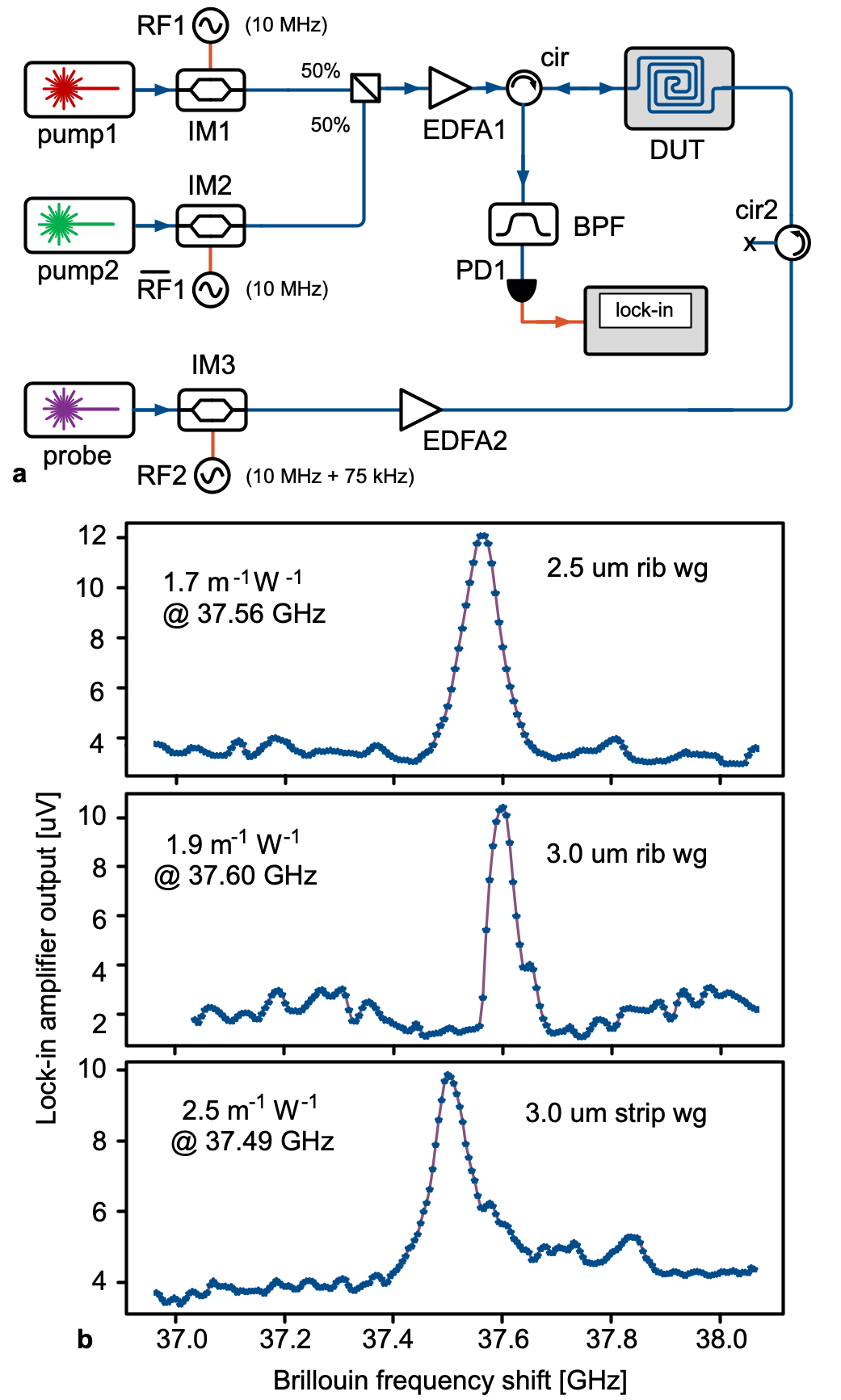}
\caption{\textbf{a} Schematic of the triple intensity modulated pump-probe lock-in amplifier setup. \textbf{b} Measured SBS signals from the 2.5~$\mu$m-wide rib waveguide, 3.0~$\mu$m-wide rib waveguide, and 3.0~$\mu$m-wide strip waveguide. All waveguides under test are 1.3~cm long straight waveguides. \textbf{c} Simulated SBS responses of the thick-SOI waveguides.}
\label{fig4}
\end{figure}

Our samples were fabricated using the standard thick-SOI platform from VTT \cite{Aalto2019}. Waveguides were defined on 3~$\mu$m thick silicon layer using stepper lithography, followed by inductively coupled plasma reactive ion etching (ICP-RIE). After etching, the wafers were annealed in a hydrogen atmosphere to reduce sidewall roughness. Lastly, a 0.5~$\mu$m thick silica cladding was deposited using low-pressure chemical vapor deposition (LPCVD) \cite{cleo24}. 

Two types of waveguides are available in this thick-SOI platform: the single-mode rib waveguide and the multi-mode strip waveguide, with cross sections shown in Fig.~\ref{fig1}\textbf{a}. The combination of these waveguides, along with the adiabatic coupler between them, enables creation of large-scale, single-mode photonic integrated circuits with high-density integration, as demonstrated in Fig.~\ref{fig3}\textbf{a}. Moreover, benefiting from the large mode area in these waveguides, nonlinear loss due to the two-photon absorption and free-carrier absorption effects is significantly reduced compared to silicon nanowires \cite{Morrison2016}.

Another big advantage of the thick-SOI platform is its ultra-low propagation loss. As illustrated in Fig.~\ref{fig1}\textbf{b} and Fig.~\ref{fig1}\textbf{c}, the electric field in thick-SOI waveguides is better confined, resulting in reduced interaction with the waveguide sidewalls. The hydrogen annealing step adopted in the thick-SOI technology further reduces sidewall roughness, contributing to lower optical losses. Fig.~\ref{fig3}\textbf{b} shows the resonance characterization of an add-drop ring resonator with a free spectral range of 5.1~GHz. The measured linewidth of 31.2 MHz corresponds to a propagation loss as low as 6.7 dB/m, which is an order of magnitude lower than the loss in the state-of-the-art silicon nanowires \cite{Rahim2019}. This propagation loss can be further reduced to 2.9 dB/m, as reported in \cite{Marin2023}.

\section{SBS characterization}
The SBS responses of waveguides in the thick-SOI platform were characterized using a triple-intensity modulated pump-probe lock-in amplifier setup, as shown in Fig.~\ref{fig4}\textbf{a}. In this setup, the intensity-modulated pump1 (1560~nm) and the probe, are coupled into the waveguide from opposite directions. To compensate the power fluctuation induced by the intensity modulation, an auxiliary pump2 (1561~nm), modulated at the same RF frequency as pump1 but with an opposite phase, is combined with the pump1. By sweeping the probe frequency across the Brillouin frequency shift range of both the waveguide and the fiber, the SBS responses of the waveguides are extracted from the lock-in amplifier. A more detailed explanation of the experimental setup can be found in \cite{Ye2024}.

Fig.~\ref{fig4}\textbf{b} presents the measured SBS responses of the rib and strip waveguides in the thick-SOI platform. Both types of waveguide exhibit a clear SBS signal near 37.6~GHz. By comparing the SBS peak from the waveguide with those from the fiber, the Brillouin gain coefficients are calculated. For the 2.5~$\mu$m-wide rib waveguide, the Brillouin gain coefficient $g_B$ is 1.7 m$^{-1}$W$^{-1}$ with a linewidth of 70~MHz. This value increases to 1.9~m$^{-1}$W$^{-1}$ with a linewidth of 57.5~MHz for the 3.0~$\mu$m-wide rib waveguide. For the 3.0~$\mu$m-wide strip waveguide, the peak Brillouin gain coefficient is 2.5~m$^{-1}$W$^{-1}$ with a linewidth of 81.5~MHz. The measured Brillouin frequency shifts match well with the SBS responses predicted from simulations, shown in Fig.~\ref{fig2}\textbf{d}. The discrepancy in the Brillouin gain coefficients between the measurements and simulations is primarily due to the polarization fluctuations during the measurements.


\begin{figure}[t!]
\centering
\includegraphics[width=\linewidth]{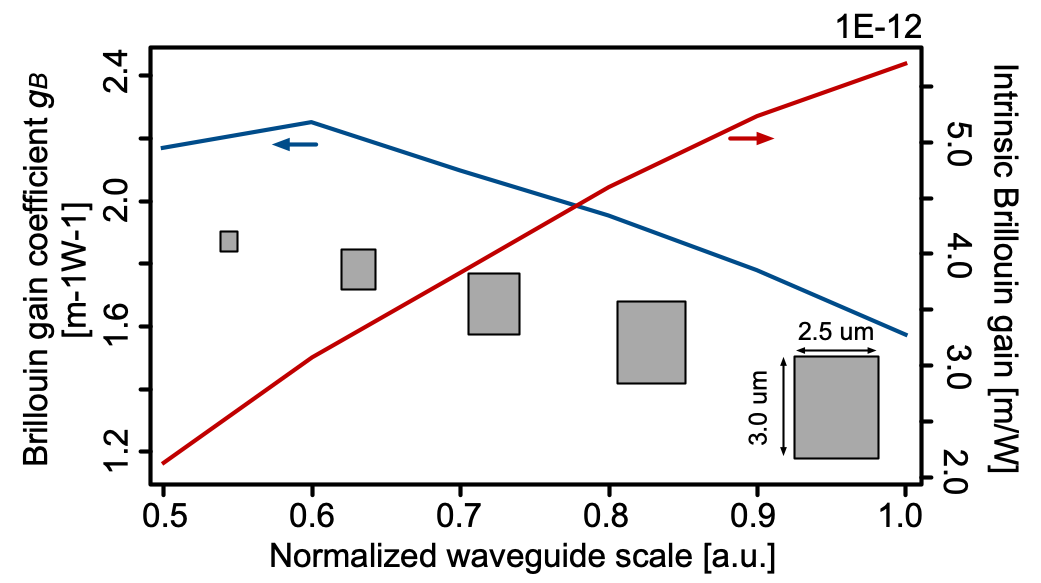}
\caption{Optimization of the thick-SOI waveguide dimension for high Brillouin gain coefficient.}
\label{fig5}
\end{figure}

\section{Discussions}
In conclusion, we have demonstrated the first observation of SBS signals in a non-suspended ultra-low-loss thick-SOI platform. To further enhance the Brillouin gain coefficient in this platform, waveguide dimensions can be optimized to maintain coupling to the leaky acoustic modes while reducing the effective mode area. Fig.~\ref{fig5} presents simulation results of the Brillouin gain coefficients as a function of the cross-section size of a strip waveguide. Initially, as the waveguide size decreases, the Brillouin gain coefficient increases due to the reduced mode area. However, beyond a certain threshold, it starts to decrease because the weaker acoustic wave confinement becomes the limiting factor. This trend is further illustrated by the intrinsic Brillouin gain, which excludes the influence of the effective mode area and continues to decrease as the waveguide shrinks. The trade-off between the acoustic wave confinement and the mode area can be further optimized leveraging techniques like genetic algorithms, as demonstrated in \cite{Lei2024, Botter2022}.

With the enhanced Brillouin gain coefficient and the low optical loss, stimulated Brillouin lasers can potentially be achieved in this platform. Additionally, the high Brillouin frequency shift makes the thick-SOI platform a promising candidate for generating pure millimeter-wave signals through SBS phonon lasing process. 

\section*{Author Contribution}
K.Y. and D.M. proposed the concept and designed the experimental plan. Y.E.M., M.C., and T.A. designed and fabricated the thick-SOI waveguide samples. K.Y. and A.K. performed the experiments with input from D.M. K.Y. and A.K. wrote the manuscript with input from all authors. D.M. supervised the project.
\label{sec:four}

\begin{acknowledgments}
The authors acknowledge funding from the European Research Council Consolidator Grant (101043229 TRIFFIC), Nederlandse Organisatie voor Wetenschappelijk Onderzoek (NWO) Start Up (740.018.021).
\end{acknowledgments}



\bibliographystyle{IEEEtran}
\bibliography{library}
\end{document}